\documentclass{mcma_v02}

\usepackage{algorithm,algpseudocode}

\title{Geometric Allocation Approach for\\
  Transition Kernel of Markov Chain}
\headlinetitle{Geometric Allocation Approach for Kernel of Markov Chain}

\authorone{Hidemaro Suwa}
\addressone{Department of Applied Physics, University of Tokyo, Tokyo 113-8656}
\countryone{Japan}
\emailone{suwamaro@looper.t.u-tokyo.ac.jp}

\authortwo{Synge Todo}
\addresstwo{Institute for Solid State Physics, University of Tokyo, Kashiwa 277-8581}
\countrytwo{Japan}
\emailtwo{wistaria@issp.u-tokyo.ac.jp}

\researchsupported{This research is supported by Grant-in-Aid for
  Scientific Research Program (Nos.\,20540364, 23540438) from JSPS and
  by Grand Challenges in Next-Generation Integrated Nanoscience,
  Next-Generation Supercomputer Project from MEXT, Japan.}

\abstract{We introduce a new geometric approach that constructs a
  transition kernel of Markov chain.  Our method always minimizes
  the average rejection rate and even reduce it to zero in many
  relevant cases, which cannot be achieved by conventional methods,
  such as the Metropolis-Hastings algorithm or the heat bath
  algorithm (Gibbs sampler).  Moreover, the geometric approach
  makes it possible to find not only a reversible but also an
  irreversible solution of rejection-free transition probabilities.
  This is the first versatile method that can construct an
  irreversible transition kernel in general cases.  We demonstrate
  that the autocorrelation time (asymptotic variance) of the 8-state Potts
  model becomes more than 14 times as short as that by the
  conventional Metropolis-Hastings algorithm.  Our algorithms are
  applicable to almost all kinds of Markov chain Monte Carlo methods
  and will improve the efficiency.
}

                                        

\acknowledgments{Most simulations were performed on T2K Supercomputer
  at University of Tsukuba.  The program was developed based on the
  ALPS library~\cite{ALPS2011s}.}

\firstpage{1}                                %
\pubyear{}                                   %
\volume{}                                    %
\issue{}                                     %
\doi{}                                       %
\communicated{}                              %
\received{}                                  %
\revised{}                                   %

\begin{document}
\section{Introduction}
The Markov chain Monte Carlo (MCMC) method, which is based on the
importance sampling and a powerful tool especially for systems with
multiple degrees of freedom, is being applied extensively across the
various disciplines, such as statistics, physics, chemistry,
bioinformatics, economics, and so
on~\cite{LandauB2005,RobertC2004}. Although an MCMC method satisfying
appropriate conditions (ergodicity) guarantees that estimators
asymptotically converge in principle~\cite{MeynT1993}, rapid
convergence is essential for the method to work in practice. In the
Monte Carlo method, if the central limit theorem holds, the variance
of expectations decreases as $\sigma^2/n$, where $n$ is the number of
samples. Then, what we have to concern is to reduce the asymptotic
variance $\sigma^2$. Since the autocorrelation of a Markov chain
exactly corresponds to the asymptotic variance, it is clearly
important to develop an update method that achieves shorter autocorrelation
time.

There are three key points for the MCMC method to be effective.  One
is the choice of the ensemble. From the view of this respect, the
extended ensemble methods, such as the multicanonical
method~\cite{BergN1992} and the replica exchange
method~\cite{HukushimaN1996}, have been proposed and applied
successfully to protein folding problems, spin glasses, etc. The
second is the selection of candidate configurations. The cluster
algorithms, e.g., the Swendsen-Wang algorithm~\cite{SwendsenW1987} and
the loop algorithm~\cite{EvertzLM1993}, can overcome the critical
slowing down by taking advantage of mapping to graph configurations in
many physical models. The third is the determination of the transition
probability, given candidate configurations. We focus our interest on
this optimization problem of the probabilities through this paper.

In the MCMC method, the (total) balance, that is, the invariance of
target distribution, is usually imposed to the transition kernel
although a kind of adaptive procedure catches much attention these
days~\cite{AndrieuT2008}.  For the optimization of the transition
probabilities, it is a guiding principle to minimize rejection rate,
the probability that a configuration stays still at the previous
state~\cite{Peskun1973}. In most practical implementations, the
Metropolis-Hastings algorithm~\cite{MetropolisRRTT1953,Hastings1970}
(we call it simply the Metropolis algorithm below) or the heat bath
algorithm~\cite{Barker1965}, namely, the Gibbs
sampler~\cite{GemanG1984}, have been used for the determination of the
transition probabilities. These canonical algorithms satisfy the
detailed balance, the reversibility, which is a sufficient condition
for the total balance. Under this condition, thanks to the simple
property that every elementary transition balances with a
corresponding inverse process, it becomes easy to find a qualified
transition probability by solving the equation for each pair of
configurations. Thus, attempts to reduce autocorrelation in the
optimization problem have concentrated within this sufficient
condition so far~\cite{Liu1996,PolletRVH2004}. However, all the
previous methods fail to minimize the rejection rate in most cases.

In this paper, we introduce a new method that constructs a transition
kernel by a geometric approach. This method can find solutions by
applying a graphical procedure, \emph{weight allocation}, instead of
solving the detailed balance equation algebraically as before.
Surprisingly, it is \emph{always} possible to find a solution that
minimizes the average rejection rate. In the meantime, it has long
been considered difficult to satisfy the total balance without
imposing the detailed balance.
This reversibility is sufficient, however, \emph{not necessary} for the
invariance of target distribution. If it is possible to find a
solution beyond the sufficient condition, further optimization can be
achieved. Our approach is the first method that can generally satisfy
the total balance without the detailed balance.  We will introduce our
geometric picture for the optimization problem and then explain
concrete algorithms for constructing a reversible and an irreversible
kernel~\cite{SuwaT2010}.  We will demonstrate its effectiveness in a
basic physical example, using the single spin update of the
ferromagnetic Potts model.
\section{Geometric Approach}
In the MCMC method, we update configuration (or state) variables
locally and run over the whole system. Now, let us consider updating
one discrete variable as an elementary process, e.g., flipping a
single spin in the Ising or Potts models~\cite{Wu1982}. Given an
environmental configuration, we would have $n$ candidates (including
the current one) for the next configuration. The weight of each
candidate configuration (or state) is given by $w_i$ ($i=1,\/ ...,
n$), to which the equilibrium probability measure is proportional.
Although the total and detailed balance are usually expressed in terms
of the weights $\{w_i\}$ and the transition probabilities $\{p_{i
  \rightarrow j}\}$ from state $i$ to $j$, it is more convenient to
introduce a quantity $v_{ij} := w_i p_{i \rightarrow j}$, which
corresponds to the amount of (raw) stochastic flow from state $i$ to
$j$.  The law of probability conservation and the total balance are
then expressed as
\begin{eqnarray}
  w_{i} & = \sum_{j=1}^n v_{ij} \qquad \forall \, i
  \label{eqn:conservation}
  \\
  w_{j} &= \sum_{i=1}^n v_{ij} \qquad \forall \, j,
  \label{eqn:bc}
\end{eqnarray}
respectively.  The average rejection rate is written as $\sum_i v_{ii}
/ \sum_i w_i$.  Also, it is straightforward to confirm that
$\{v_{ij}\}$ satisfy $ v_{ij} = \min[ w_i, w_j ] / (n-1) \ ( i \neq j ) $
for the Metropolis algorithm with the flat proposal distribution, and
$  v_{ij} = w_i w_j / \sum_{k=1}^n w_k \ ( \forall \, i, \, j )$
for the heat bath algorithm (Gibbs sampler), where the detailed
balance, i.e., the absence of net stochastic flow, is manifested by
the symmetry under the interchange of the indices:
\begin{equation}
  v_{ij} = v_{ji} \qquad \forall \, i, \, j.
\label{eqn:dbc}
\end{equation}
\begin{figure}
\begin{center}
\includegraphics[width=5.9cm]{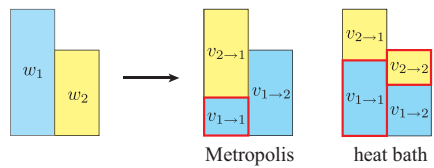}
\caption{\small{Example of the weight allocation by the Metropolis and heat bath algorithms
  for $n=2$. The regions with thick frame denote the rejection rates.}}
\label{fig:landfill}
\end{center}
\end{figure}
Our task is to find a set $\{v_{ij}\}$ that minimizes the average
rejection rate while satisfying Eqs.~(\ref{eqn:conservation}) and
(\ref{eqn:bc}).  The procedure for the task can be understood visually
as \emph{weight allocation}, where we move (or allocate) some amount of
weight ($v_{ij}$) from state $i$ to $j$ keeping the entire shape of
the weight boxes intact.  For catching on this allocation picture, let
us think at first the case with $n=2$ as in the single spin update of
the Ising model.  Fig.~\ref{fig:landfill} shows the allocation when
the Metropolis and heat bath algorithms are applied, where the average
rejection rate ($\propto v_{11} + v_{22}$) clearly remains finite.
Indeed, for $n=2$ the Metropolis algorithm gives the best solution,
i.e., the minimum average rejection rate even within the total balance
[see Eq.~(\ref{eqn:rejection}) below].

For $n \geq 3$, these two methods fail to minimize the rejection rate
as we will mention. Besides, a generic method that accomplishes the
minimization has not been known before. We will show that we can
easily make it possible by this geometric picture. Although many
optimal solutions are found actually, here we will introduce two
specific algorithms. One makes a reversible kernel, and the other
makes an \emph{irreversible} kernel without the detailed balance.
\subsection{Reversible Kernel}
For describing our algorithm, let us introduce an operation named Swap:
\begin{algorithmic}
\State Swap( $i$, $j$, $w$ ) \{
\State \hspace{3mm} $v_{ii} \gets v_{ii} - w$
\State \hspace{3mm} $v_{ij} \gets v_{ij} + w$
\State \hspace{3mm} $v_{ji} \gets v_{ji} + w$
\State \hspace{3mm} $v_{jj} \gets v_{jj} - w$
\State \}.
\end{algorithmic}
We note that if $\{v_{ij}\}$ satisfy the conditions
(\ref{eqn:conservation}), (\ref{eqn:bc}) and (\ref{eqn:dbc}), this
Swap operation does not break them.  A certain algorithm for the
construction of reversible kernel that minimizes the average rejection
rate is described in Algorithm~\ref{alg:reversible}.
\begin{algorithm}[t!]
\caption{Construction of Reversible Kernel with Minimized Rejection}
\begin{algorithmic}
  \\
  \State Sort $n$ candidate configurations as $w_1 \geq w_2 \geq w_3 \geq ... \geq w_n$ ($ n \geq 3 $).
  \\
\State $v_{ij} \gets w_i \/ \delta_{ij}$
  \State $w_{\rm diff} \gets w_1 - w_2$
  \State $ S_3 \gets \sum_{i=3}^n w_i$
  \\
  \If{$w_{\rm diff} \geq S_3$}
  \For{$i = 2,\/ ..., n$}
  \State Swap( 1, $i$, $w_i$ ) \hspace{2.3cm}// $v_{ii}$ becomes 0
  \EndFor
  \Else
  \For{$i = 3,\/ ..., n$}
  \State $v \gets w_{\rm diff} * w_{i} / S_3$
  \State Swap( 1, $i$, $v$ )
  \EndFor  \hspace{4cm}// $v_{11} = v_{22} \geq v_{33} \geq ... \geq v_{nn}$
  \For{$j = n,\/ ..., 2$}
  \State $v' \gets v_{jj} / ( j - 1 )$
  \For{$k = j-1,\/ ..., 1$}
  \State Swap( $j$, $k$, $v'$ )
  \EndFor  \hspace{3.5cm}// $v_{11} = v_{22} \geq ... \geq v_{j-1,j-1}$ and $v_{jj} = 0$
  \EndFor
  \EndIf
\end{algorithmic}
\label{alg:reversible}
\end{algorithm}
This algorithm starts with the diagonal matrix $[ v_{ij} ]$ and uses
only Swap operation for construction. Therefore the three conditions
(\ref{eqn:conservation}), (\ref{eqn:bc}) and (\ref{eqn:dbc}) are
automatically satisfied in the whole procedure.  This algorithm can be
depicted visually as Algorithm~1 in Fig.~\ref{fig:landfill-2}. As a
result, the self-allocated weight that produces rejection is expressed
as
\begin{eqnarray}
v_{ii} = \left\{ \begin{array}{lll} \max( 0, \, w_1 - \sum_{i=2}^n w_i ) & \qquad i = 1  \\
    0 \hspace{2mm} & \qquad i \geq 2
  \end{array} \right.
\label{eqn:rejection}
\end{eqnarray}
That is, a rejection-free solution can be obtained, if
\begin{eqnarray}
w_1 \leq \frac{S_n}{2} \equiv \frac{1}{2}\sum_{k=1}^n w_k
\label{eq:rejection-free}
\end{eqnarray}
is satisfied.  In contrast, when inequality
(\ref{eq:rejection-free}) is not satisfied, one has to necessarily
assign the maximum weight to itself since it is larger than the sum of
the rest.  Thus, the present solution is optimal in the sense that it
minimizes the average rejection rate.
\subsection{Irreversible Kernel}
\begin{figure}[t]
\begin{center}
\includegraphics[width=12cm]{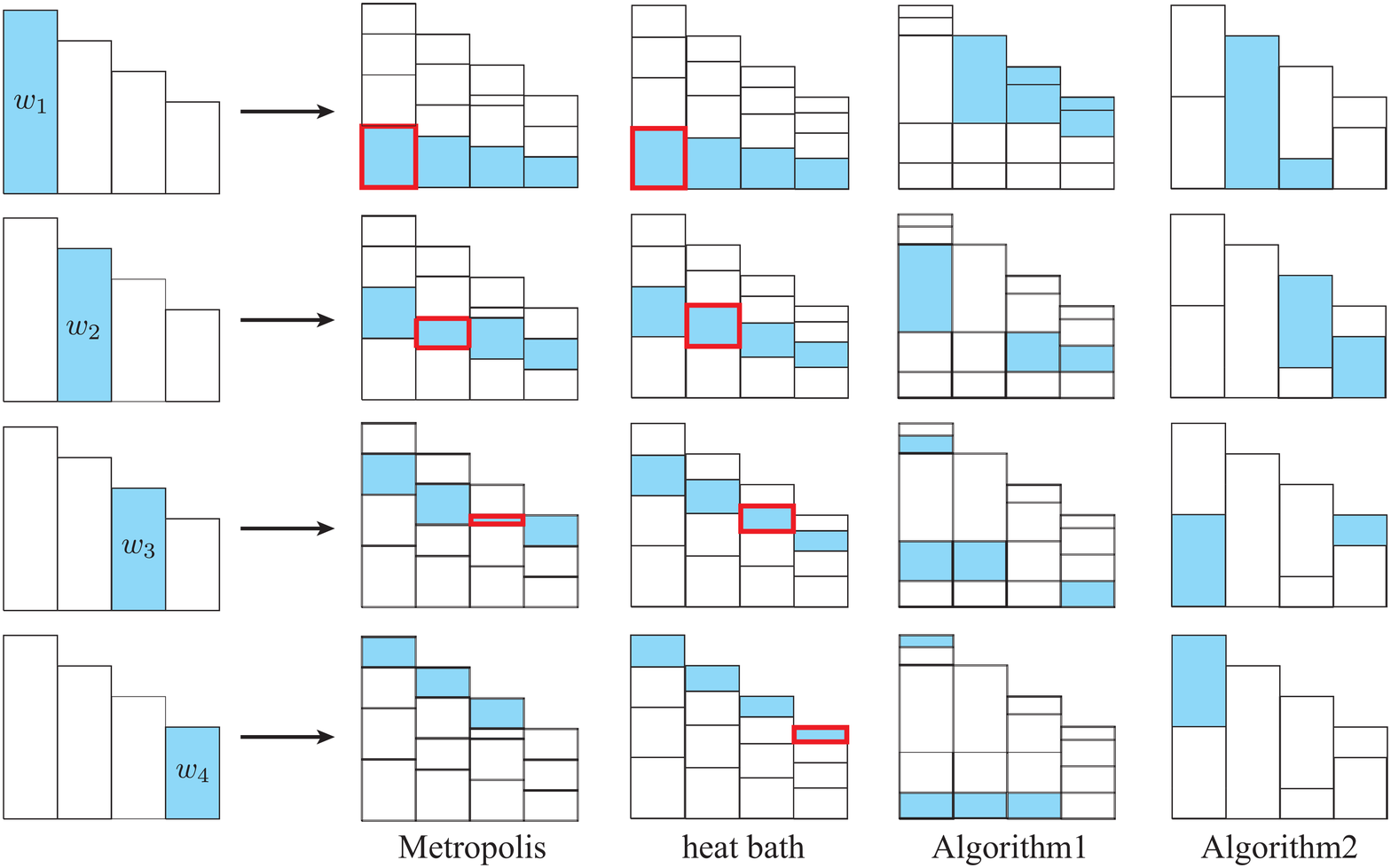}
\caption{\small{Example of weight allocation by the Metropolis, the heat
  bath, and the proposed two algorithms for $n=4$. Algorithm~1
  constructs a reversible kernel, and Algorithm~2 does an irreversible
  kernel. Both proposed algorithms minimize the average rejection rate
  in general, and they are rejection free in this case while the
  conventional methods remain finite rejection rates as
  indicated by the thick frames.}}
\label{fig:landfill-2}
\end{center}
\end{figure}
Next, we show another algorithm that constructs an irreversible
kernel~\cite{SuwaT2010}. The whole algorithm is described in
Algorithm~\ref{alg:irreversible}.
\begin{algorithm}
\caption{Construction of Irreversible Kernel with Minimized Rejection}
\begin{algorithmic}
  \\
  \State Choose a configuration that has the maximum weight and number it 1.
  \State Sort other configurations in an arbitrary order.
  \State $i \gets 1$ \State $j \gets 2$ \While{ $ i \leq n$ } \State
  $w_r \gets w_i$ \While{ $w_r > 0$ } \If{ $w_r \geq w_j$ } \State
  $v_{ij} \gets w_j$ \State $w_r \gets w_r - w_j$ \If{ $j = n$} \State
  $j \gets 1$ \Else \State $j \gets j + 1$
  \EndIf
  \Else 
  \State $v_{ij} \gets w_r$ 
  \State $w_j \gets w_j - w_r$
  \State $w_r \gets 0$
  \EndIf
  \EndWhile
  \State $i \gets i + 1$
  \EndWhile
\end{algorithmic}
\label{alg:irreversible}
\end{algorithm}
In the algorithm, if two or more configurations have the same maximum
weight, choose one of them at first. Any order of configurations
accomplishes the same minimized rejection rate. In the above
procedure, all the boxes are filled without any space as well as the
reversible case, as in Fig.~\ref{fig:landfill-2}; it satisfies the two
conditions (\ref{eqn:conservation}) and (\ref{eqn:bc}). However, the
reversibility (\ref{eqn:dbc}) is broken.  (For example, $v_{12} > 0$,
but $v_{21}=0$ as depicted in the figure.)  Since $w_1$ is the
maximum, it is also clear that the second and subsequent boxes must be
already saturated when the allocation of its own weight is initiated.

The rejection rate becomes the same with the previous
reversible kernel as formulated in Eq.~(\ref{eqn:rejection}).  In
contrast to the reversible case, a net stochastic flow is introduced
as the result of breaking the detailed balance, and it is expected to
further boost up the sampling efficiency~\cite{DiaconisHN2000}.
\section{Benchmark test}
\begin{figure}
\begin{center}
\includegraphics[width=7.0cm]{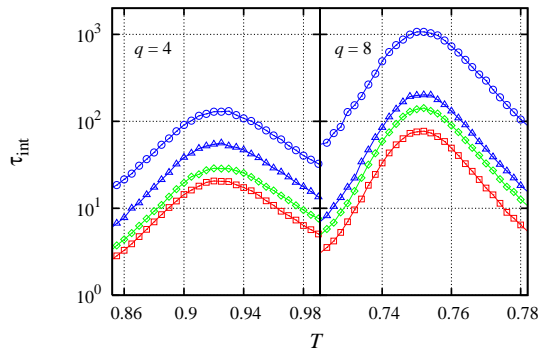}
\caption{\small{Autocorrelation time of the square of order parameter near the transition temperature ($T \simeq 0.910$ and 0.745, respectively) in the 4-state (left) and 8-state (right) Potts models by the Metropolis (circles), heat bath (triangles), LOU (diamonds), and present (squares) methods. The results of present two algorithms are the same in this scale. The system size is $16 \times 16$. 
  The error bars are the same order with the point sizes.}}
\label{fig:Potts-stf} \end{center} \end{figure}

In order to assess the effectiveness of the present algorithms, we
investigate the autocorrelations in the ferromagnetic $q$-state Potts
models on the square lattice~\cite{Wu1982}, which exhibit a continuous
($q \le 4$) or first-order ($q>4$) phase transition at $T=1 / \ln
(1+\sqrt{q})$.  We calculate the autocorrelation time of the square of
order parameter for $q=4$ and 8 by several algorithms.  The
autocorrelation time $\tau_{\rm int}$ is estimated through the
relation: $ \sigma^2 = ( 1 + 2 \tau_{\rm int} ) \sigma_{0}^2$, where
$\sigma_0^2$ and $\sigma^2$ are the variances of the estimator without
considering autocorrelation and with calculating correlation from the
binned data using a bin size much larger than the $\tau_{\rm
  int}$~\cite{LandauB2005}.  In Fig.~\ref{fig:Potts-stf}, it is
clearly seen that our algorithms significantly boosts up the
convergence in both models in comparison with the conventional
methods.  In the 4 (8)-state Potts model, the autocorrelation time
becomes nearly 6.4 (14) times as short as that by the Metropolis
algorithm, 2.7 (2.6) times as short as the heat bath algorithm, and
even 1.4 (1.8) times as short as the locally optimal update
(LOU)~\cite{PolletRVH2004}, which was considered as one of the best
solutions before our approach. The autocorrelations of our two
algorithms are much the same both for $q=4,8$. We also note that our
irreversible algorithm improves the efficiency more than 100 times as
much as that by the heat bath algorithm in a quantum spin
model~\cite{SuwaT2010}.
 \section{Conclusion}
We have introduced the new geometric approach for optimization of
transition probabilities and the two concrete algorithms that always
minimizes the average rejection rate in the MCMC method. One
constructs a reversible kernel, and the other does an irreversible
kernel, which is the first versatile method that constructs an
irreversible chain in general cases. We showed our algorithms
significantly improve the sampling efficiency in the ferromagnetic
Potts models. The autocorrelations of our two algorithms are much the
same in the model; the net stochastic flow does not matter to the
efficiency. However, it is generally possible for the flow to play an
important role to the convergence. The introduction of efficient flow
needs to be researched in the future. Finally, we note that our
algorithm for irreversible kernel can be generally extended to
continuous variables, which will be presented in an other
report.
\bibliography{suwa}
\end{document}